\documentclass[journal]{IEEEtran}
\usepackage{tikz}
\usepackage{xcolor,soul,framed} 
\usepackage{tabularx} 
\usepackage{cite}
\usepackage[flushleft]{threeparttable}
\usepackage{url}
\usepackage{graphicx}

\colorlet{shadecolor}{yellow}
\usepackage[cmex10]{amsmath}
\usepackage{array}
\usepackage{mdwmath}
\usepackage{mdwtab}
\usepackage{eqparbox}
\usepackage{url}
\usepackage{todonotes}
\usepackage[figure]{algorithm2e} 
\usepackage{listings}    
\usepackage{subfig} 
\usepackage{hyperref}
\usepackage{etoolbox}

\usepackage{tabularx}
\usepackage{adjustbox}

\makeatletter
\patchcmd{\@verbatim}
  {\verbatim@font}
  {\verbatim@font\small}
  {}{}
\makeatother

\begin{document}
\bstctlcite{IEEEexample:BSTcontrol}
\title{Synthesis of Parallel Synchronous Software\\\vspace*{-.2cm}}
\author{Pantea~Kiaei,~\IEEEmembership{Student~Member,~IEEE, }
        Patrick~Schaumont,~\IEEEmembership{Senior~Member,~IEEE}
        \\\vspace*{-1cm}
\IEEEcompsocitemizethanks{\IEEEcompsocthanksitem 
This research was supported in part by NIST Award 70NANB17H280 and NSF Award 1931639.\protect\\
The authors were with the Department of Electrical and Computer Engineering, Virginia Tech, Blacksburg, VA 24061 USA. (E-mail: pantea95@vt.edu; schaum@vt.edu)
\vspace{-.5cm}
\IEEEcompsocthanksitem 
}
}

\markboth{IEEE EMBEDDED SYSTEMS LETTERS
}{Authors}{}

\maketitle

\begin{abstract}
In typical embedded applications, the precise execution time of the program does not matter, and it is sufficient to meet a real-time deadline. However, modern applications in information security have become much more time-sensitive, due to the risk of timing side-channel leakage. The timing of such programs needs to be data-independent and precise.
We describe a parallel synchronous software model, which executes as N parallel threads on a processor with word-length N. Each thread is a single-bit synchronous machine with precise, contention-free timing, while each of the N threads still executes as an independent machine. The resulting software supports fine-grained parallel execution. In contrast to earlier work to obtain precise and repeatable timing in software, our solution does not require modifications to the processor architecture nor specialized instruction scheduling techniques. In addition, all threads run in parallel and without contention, which eliminates the problem of thread scheduling.
We use hardware (HDL) semantics to describe a thread as a single-bit synchronous machine. Using logic synthesis and code generation, we derive a parallel synchronous implementation of this design. We illustrate the synchronous parallel programming model with practical examples from cryptography and other applications with precise timing requirements.
\vspace{-.2cm}
\end{abstract}


%
\IEEEpeerreviewmaketitle


     \vspace{-.8cm}

\section{Introduction}

Producing software with precise, repeatable timing is a challenging task. First, the software application itself may have data-dependent processing complexity, such as with data-dependent loops. Second, the execution time of the application on the processor may be affected by the memory hierarchy and the run-time state of the processor. Third, the timing of the execution may be affected by resource contention when several parallel threads share the same processor resource. Among these three problems, the second and the third are most difficult because they are outside of the control of the programmer. In cryptographic applications, data-dependent timing variations may be exploited as timing side-channel leakage, either directly as an effect of data-dependent control flow, or indirectly as an effect of contention on shared processor resources. To avoid timing side-channels, we need data-independent timing. 

In this contribution, we propose a programming model that yields these  timing characteristics. We contrast our proposal with earlier work towards precise software timing for embedded applications, PRET \cite{DBLP:conf/rtss/LeeRZ17, DBLP:conf/rtas/ZimmerBSL14}. A fundamental idea of PRET is to use instruction scheduling to avoid resource contention in the processor in the pipeline. By spacing the instructions of timing-critical threads several cycles apart, stall-free execution is achieved in the pipeline. As a consequence, the timing of individual threads is repeatable regardless of the processor state. To ensure overall processor utilization, PRET combines multiple timing-critical threads with time-interleaving and a customized instruction scheduling technique \cite{DBLP:conf/rtas/ZimmerBSL14}.

Our insight in this paper is that such time-sensitive threads can also be combined {\em spatially} within a processor word, instead of {\em temporally} using interleaved instruction streams. The advantage of spatially combining the threads (instead of using time-based interleaving) is that we don't need to adapt the processor for interleaved instruction execution. To implement a spatial arrangement of threads, we organize each thread as a single-bit program, and execute the overall application as a vectorized version of the single-bit program. We emphasize that the proposed model goes beyond software bit-slicing \cite{biham1997fast}, which is strictly functional and ignores the control flow and the state within each slice.

To simplify the development of single-bit programs, we adopt a synchronous execution model. A single-bit program is captured as a synchronous Finite State Machine with Datapath (FSMD), and the execution of this program follows a sequential schedule of the bit-operations that define the FSMD. The vectorized form of the single-bit program is then achieved with bitwise instructions over the processor word. The vectorized form is a {\em parallel} synchronous program. Since each thread has its own state, each thread executes as an independent FSMD. However, the instruction count for one iteration of the overall program is constant and repeatable, and therefore the execution time of these FSMD threads becomes repeatable too. A prototype implementation of a synthesis tool starts from a Verilog input specification and generates C code with inline assembly optimized for an embedded target. We demonstrate several useful examples of parallel synchronous programming (PSP).

The outline of this contribution is as follows. Section~\ref{sec:prelims} develops the cardinal components of parallel synchronous programming. Section~\ref{sec:parallelSynchronousSoftware} discusses an example and proposes a code generation methodology. Section~\ref{sec:experiments} describes experimental results. Section~\ref{sec:conclusion} concludes the paper.

     \vspace{-.5cm}

\section{Preliminaries}\label{sec:prelims}
We develop a software execution model that leads to repeatable, data-independent timing. We first define what is meant by repeatable and data-independent timing in software. We then describe software bitslicing, which can offer such timing characteristics for functions (i.e. straight-line stateless programs). Next, we explain how to extend the semantics of software bitslicing from straight-line programs to synchronous FSMD. The result is a Parallel Synchronous Program (PSP).

\vspace{-.4cm}

\subsection{Desired Timing Properties} 
Programs written as PSP aim for repeatable timing as well as data-independent timing. The former is useful in real-time embedded software design, while the latter is useful for secure systems design. We motivate and differentiate each property.

Edwards {\em et al.} make a distinction between repeatable timing and predictable timing \cite{DBLP:conf/iccd/EdwardsKLLPS09}. Repeatable timing means that every correct execution of a program uses the same timing. Repeatable timing is desired as a property of the program, not of the program running on a specific processor.
Repeatable timing is needed in the context of real-time applications when timing jitter is a concern. For example, when a physical sensor must be read from software at a specific sample rate, then the software needs to have repeatable timing. Jitter is typically caused by resource contention and interrupts. 

A second relevant domain for PSP is that of secure software. In recent years, a rich collection of attacks have been found to exploit the {\em implementation} characteristics of secure software rather than the program logic itself. The best known of these are side-channel attacks and micro-architectural attacks, which rely on precise execution time measurement \cite{DBLP:journals/jce/GeYCH18}. To thwart these attacks, software with (secret-)data-independent timing is needed. This is hard because modern micro-architectures are rife with architectural contention and context-dependent timing. Even if there are no obvious dependencies in the program logic, there may still be hidden dependencies in the micro-architecture. The cryptographic community is well aware of the risk of timing-based side-channel leakage, leading to the design of so-called {\em constant-time} software that avoids data-dependencies in the program execution time \cite{DBLP:conf/uss/AlmeidaBBDE16}. The resulting programs are not literally constant-time, but rather they adopt data-independent control flow and memory access patterns.

We argue that software written as a Parallel Synchronous Program (PSP) provides repeatable timing as well as data-independent timing. PSP achieves these properties by combining two concepts: software bitslicing and synchronous FSMD. The following subsections introduce both.


\vspace{-.5cm}
\subsection{Bitslicing} 
Software bitslicing was originally proposed for high throughput software implementations \cite{biham1997fast}. In this model of programming, a program is expanded into 1-bit (Boolean) operations as follows. A $k-$bit variable with bits $b_{k-1} ... b_1 b_0$ is distributed over $k$ registers $R_{k-1} ... R_1 R_0$, such that register $R_i$ holds bit $b_i$. An $N-$bit processor operates as an $N-$way SIMD processor, processing $N$ instances of the $k-$bit variable in parallel, and storing these instances in $k$ registers. Bitsliced programs are Boolean programs written with bit-wise logic operations. The rationale of bitslicing is that it guarantees full utilization of the processor word-length. The absence of control flow ensures that each iteration through a bitslice function uses the same amount of instructions. In addition, the absence of state (memory) in a bitslice function eliminates cache timing effects. For this reason, bitslicing is often applied in the context of developing programs that are {\em constant-time} (in the cryptographic sense). However, software bitslicing is insufficient as a general-purpose methodology for software. Because bitslice functions do not have control flow, control operations are typically emulated using non-bitsliced logic surrounding bitsliced expressions. This prevents individual slices from operating as independent threads of control. Bitslice programming essentially applies only to functions. The management of the program state resides outside of the bitslice logic.

\vspace{-.5cm}

\begin{figure}[t]
    \centering
    \includegraphics[width=.8\linewidth]{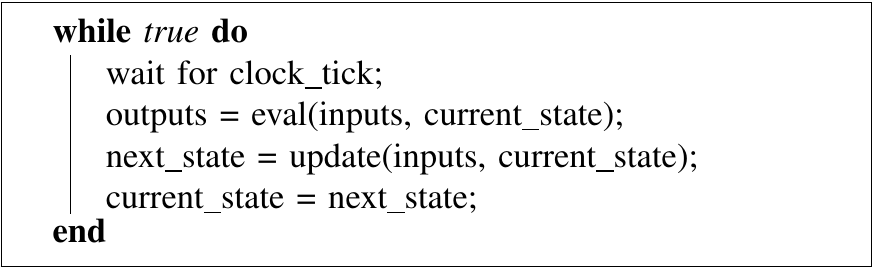}
    \vspace{-.1cm}
 \caption{Basic structure of a synchronous program}
 \label{fig:syncProg}
      \vspace{-.7cm}
\end{figure}

\subsection{Synchronous FSMD} 
We next describe how to introduce control flow and state into software bitslicing. A Boolean (1-bit) program does not offer the concept of an address space or control flow instructions. Therefore, we propose introducing the control flow through the intermediary of a synchronous FSMD. \mbox{FSMDs} are common in digital hardware design, and they are routinely applied in register-transfer-level designs. An FSMD is a synchronous model of computation combining a datapath and a finite-state controller. Computations are done on the datapath under control of the FSM \cite{schaumont2012practical}. Each synchronous clock cycle, the FSM computes a single state transition and selects one or more operations in the datapath. The execution of datapath operations depends on the current state of the FSM, and the state transition conditions in the FSM depend on the current state of the datapath. Conditional control flow is expressed using dataflow-like semantics: the datapath will compute both the true and false case of the control condition, and the correct result will be selected using multiplexing.

To map a synchronous FSMD into software, we adopt a synchronous execution model as shown in Figure \ref{fig:syncProg}. Every loop in this program corresponds to a single clock cycle of the synchronous FSMD model. The software awaits the occurrence of a clock tick to read all inputs and evaluate all outputs concurrently \cite{caspi2007synchronous}. The {\tt eval()} function in Figure \ref{fig:syncProg} computes the FSM next-state as well as the datapath next-state.  The {\tt update()} function adjusts the current state of the FSM and the datapath to the next. State update is handled synchronously: each state variable in the synchronous FSMD is split into two copies, the {\em current\_state} and the {\em next\_state}. This avoids race conditions, and ensures that the program will always compute the same result regardless of the scheduling of {\tt eval()} and {\tt update()}. The PSP is implemented as N parallel copies of the program of Figure \ref{fig:syncProg}, where every thread is tied to the same global clock tick, and each thread is a 1-bit program expressed as a synchronous FSMD.

\vspace{-.3cm}
\section{Synthesis of Parallel Synchronous Software}
\label{sec:parallelSynchronousSoftware}
We next demonstrate how to create PSP. We first describe the example PSP design of a parallel greatest common divisor program, and next discuss a design flow that synthesizes PSP software from a synchronous FSMD description.

\begin{figure}[t]
    \centering
    \includegraphics[width=.8\linewidth ]{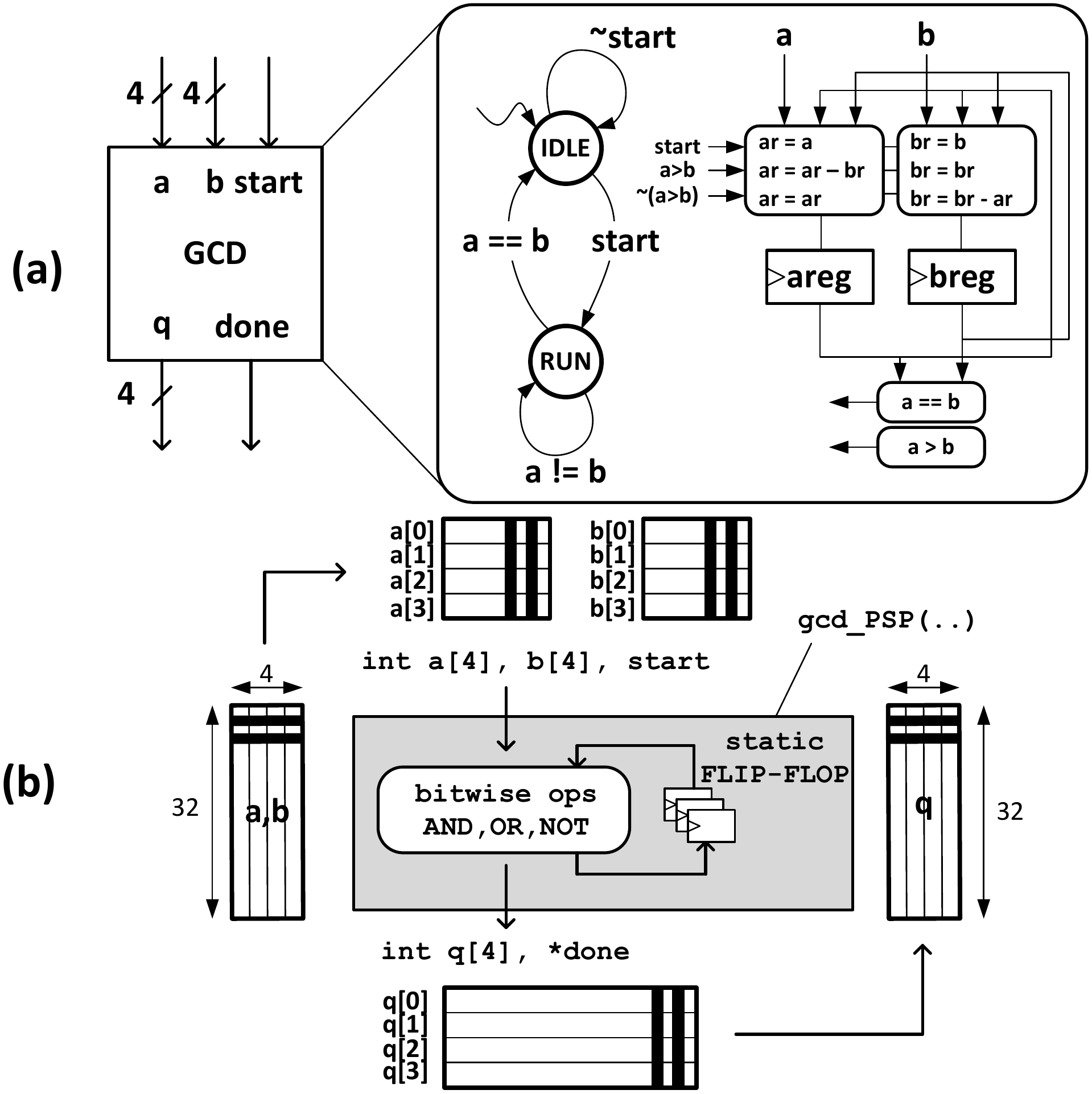}
    \caption{GCD (a) Interface and (b) FSMD model}
    \label{fig:gcd}
     \vspace{-.8cm}
\end{figure}

\paragraph{Example} Figure \ref{fig:gcd}a shows the outline of a 4-bit GCD module. We express the functionality of the GCD algorithm as an FSMD model. After a {\tt start} control pulse, the module reads two 4-bit inputs {\tt a} and {\tt b}, and repeatedly subtracts the smaller value from the larger value until they are equal. A {\tt done} pulse is generated to indicate completion of the algorithm. A two-state control FSM drives the loading of two 4-bit registers {\tt a} and {\tt b} and their iterative computation.

A PSP version of the GCD algorithm for a 32-bit processor executes 32 parallel copies of the GCD. We create this software by converting the FSMD to a gate-level netlist using logic synthesis. We target a generic technology with a logically complete set of primitive functions (such as {\tt AND}, {\tt OR} and {\tt NOT}) as well as a storage element such as a flip-flop (Figure \ref{fig:gcd}b). The outcome of the logic synthesis is a netlist in terms of logic elements. 
We then rewrite the netlist as a sequential function by leveling the netlist according to data dependencies from input to output. The logic cells are replaced by bit-wise operations, and the flip-flops are replaced by {\tt static} (or global) variables. The resulting function declaration is as follows.
     \vspace{-.25cm}

\begin{verbatim}
  gcd_PSP(int a[4], int b[4], // data input
          int q[4],           // data output
          int start,          // control in
          int* done);         // status out
\end{verbatim}
     \vspace{-.2cm}

Each invocation of this function corresponds to a single synchronous iteration (one clock cycle of the synchronous FSMD). An important difference between the circuit of Figure \ref{fig:gcd}a and the PSP function in Figure \ref{fig:gcd}b is the degree of parallelism; The circuit in Figure \ref{fig:gcd} computes a single GCD whereas the {\tt gcd\_PSP} function is a software design that computes 32 {\em concurrent} GCD algorithms independently, each with their own {\tt start} and {\tt done} bits. The inputs and outputs of {\tt gdc\_PSP} are in bitsliced form. For example, {\tt a[2]} contains the second bit of 32 different inputs. Hence, a call to {\tt gcd\_PSP} needs to transpose the input and output arguments.

\paragraph{An Automated Flow}\label{par:automatedflow} We implemented a software synthesis flow for PSP that starts from an FSMD description in a Verilog program. An open-source Verilog synthesis tool \cite{yosys} converts the FSMD into a netlist in terms of generic target technology for Boolean logic and a state element. The target library for logic synthesis is adjusted in function of the targeted processor. Table \ref{tab:instructions} demonstrates a sample mapping for several embedded processors. The state elements (flip-flop) are mapped to {\tt static} variables. 

The netlist is then converted to software as follows. The netlist is topologically sorted, from the primary inputs and flip-flop outputs to the primary outputs and the flip-flop inputs. Next, each primitive gate is converted to a bitwise operation which is either emulated in C or else added through inline assembly. We rely on the C compiler to create a sequential schedule for the gate netlist that will minimize the register pressure on the processor. The following section applies the automated flow on several examples.

\begin{table}[t]
\centering
\caption{Instructions targeted by PSP synthesis}
\vspace{-.2cm}
\begin{tabular}{l|l}
\multicolumn{1}{c|}{\textbf{processor}} & \multicolumn{1}{c}{\textbf{suitable instructions for PSP}} \\ \hline
ARM Cortex-M4                           & AND, BIC, EOR, MOV, MVN, ORN, ORR         \\
RISC-V                                  & AND, OR, XOR                              \\
MSP430                                  & AND, BIC, BIS, XOR                        \\
AVR                                     & AND, COM, EOR, OR                         
\end{tabular}
\label{tab:instructions}
    \vspace{-.2cm}

\end{table}

     \vspace{-.5cm}

\section{Experimental Results} \label{sec:experiments}

\begin{table}[]
\centering
\caption{Evaluated encryption ciphers and comparison of performance of the PSP and normal implementations of them}
\vspace{-.2cm}
\resizebox{\textwidth/2-.1in}{!}{
\begin{tabular}{l|cccc|cc|c}
                                     & \multicolumn{4}{c|}{\textbf{cipher properties}}                           & \multicolumn{2}{c|}{\textbf{\begin{tabular}[c]{@{}c@{}}speed\\ (cycles/byte)\end{tabular}}} & \multicolumn{1}{l}{}     \\
\multicolumn{1}{c|}{\textbf{cipher}} & \textbf{block size} & \textbf{key size} & \textbf{rounds} & \textbf{type} & \multicolumn{1}{l}{\textbf{PSP}}           & \multicolumn{1}{l|}{\textbf{normal}}           & \textbf{speedup}         \\ \hline
SIMON                                & 128                 & 128               & 68              & Feistel       & 744.48                                     & 1315.63                                        & 1.7$\times$ \\
PRESENT                              & 64                  & 80                & 31              & SPN           & 399.61                                     & 1069.06                                        & 2.6$\times$ \\
LED                                  & 64                  & 64                & 32              & SPN           & -                                          & -                                              & -                        \\
Midori                               & 64                  & 128               & 16              & SPN           & 236.90                                     & 2233.38                                        & 9.4$\times$
\end{tabular}}
\label{tab:ciphers}
\vspace{-.7cm}
\end{table}

\begin{table*}[t]
\vspace{-.2cm}
\caption{Evaluation of parallel synchronous examples on $48$MHz Cortex-M4F processor}
\centering
\small
\setlength\tabcolsep{2pt}
\begin{tabular}{l|ccc|cccccc|ccc|c|}
\cline{2-14}
                                       & \multicolumn{3}{c|}{\textbf{performance and cost}}                                                                                                                                                                                                                          & \multicolumn{10}{c|}{\textbf{instructions breakdown}}                                                                                                                                                                                                                                                      \\ \hline
\multicolumn{1}{|c|}{\textbf{example}} & \multicolumn{1}{c|}{\textbf{\begin{tabular}[c]{@{}c@{}}number of cycles\\ (32 parallel runs)\end{tabular}}} & \multicolumn{1}{c|}{\textbf{\begin{tabular}[c]{@{}c@{}}throughput\\ (Kbps)\end{tabular}}} & \textbf{\begin{tabular}[c]{@{}c@{}}code size\\ (Kb)\end{tabular}} & \multicolumn{1}{c|}{\textbf{AND}} & \multicolumn{1}{c|}{\textbf{ORR}} & \multicolumn{1}{c|}{\textbf{BIC}} & \multicolumn{1}{c|}{\textbf{EOR}} & \multicolumn{1}{c|}{\textbf{ORN}} & \textbf{MVN} & \multicolumn{1}{c|}{\textbf{MOV}} & \multicolumn{1}{c|}{\textbf{STR}} & \textbf{LDR} & \textbf{overhead} \\ \hline
\multicolumn{1}{|l|}{GCD}              & 382                                                                                                           & -                                                                                         & 11.88                                                             & 28                                & 34                                & 6                                 & 9                                 & 7                                 & 8            & 21                                & 28                                & 73           & 54.46\%           \\ \hline
\multicolumn{1}{|l|}{PWM}              & 239                                                                                                           & -                                                                                         & 11.82                                                             & 29                                & 20                                & 11                                & 6                                 & 2                                 & 8            & 0                                 & 23                                & 39           & 44.93\%           \\ \hline
\multicolumn{1}{|l|}{SIMON bit-parallel}            & 381,175                                                                                                     & 515.79                                                                                    & 23.40                                                             & 907                               & 470                               & 180                               & 367                               & 27                                & 4            & 18                                & 1033                              & 2002         & 60.96\%           \\ \hline
\multicolumn{1}{|l|}{SIMON bit-serial} & 15,370,190                                                                                                  & 12.79                                                                                     & 18.81                                                             & 854                               & 313                               & 23                                & 16                                & 19                                & 13           & 7                                 & 593                               & 686          & 50.95\%           \\ \hline
\multicolumn{1}{|l|}{PRESENT} & 102,301                                                                                                  & 960.93                                                                                     & 17.79                                                             & 226                               & 282                               & 60                                & 119                                & 70                                & 32           & 24                                 & 454                               & 861          & 62.92\%           \\ \hline
\multicolumn{1}{|l|}{LED} & 139,949                                                                                              &     702.43                                                                                 & 20.29
                                                             & 379                               & 301                               & 80                                & 395                                & 60                                & 60           & 138                                 & 556                               & 1258          & 60.49\%           \\ \hline
\multicolumn{1}{|l|}{Midori} & 60,646                                                                                              & 1620.95                                                                  & 18.28

                                                             & 336                               & 265                               & 60                                & 242                                & 124                                & 78           & 91                                 & 438                               & 930          & 56.90\%           \\ \hline
\end{tabular}
\label{tab:examples}
     \vspace{-.7cm}

\end{table*}


We analyze our flow and the resulting performance using several examples. We target the $48$ MHz ARM Cortex-M4F processor, which comes with the Texas Instruments MSP432P401R Launchpad and implements the ARMv7E-M architecture.
Table~\ref{tab:examples} summarizes our results. The numbers reported on this table are compiled with size optimization ({\tt -Os}).

The first two examples, GCD and PWM, illustrate the general-purpose nature of PSP as well as its real-time characteristics. For these examples, Table~\ref{tab:examples} lists the number of processor clock cycles per synchronous cycle. Computing 32 parallel GCD's thus takes 382 clock cycles per synchronous cycle, \textit{i.e.}, per iteration of the GCD while-loop. 

The Pulse Width Modulator (PWM) generates pulses with a fixed period while having different duty cycles. The PSP version of this function in a 32-bit architecture can generate 32 pulses with varying cycles of duty at the same time.  Our implementation demonstrates a PWM with 8-bit resolution. The synchronous cycle of our PWM uses 239 ARM cycles, which provides a minimum pulse width of $\frac{239}{48\text{MHz}} = 4.98\mu s$ and a period of $2^{8} \times \frac{239}{48\text{MHz}} = 1,275\mu s$ or $784\text{Hz}$.

The second group of examples are taken from cryptography \cite{beaulieu2015simon,presentcipher,guo2011led,banik2015midori}. Their characteristics are summarized in Table~\ref{tab:ciphers}.
SIMON 128/128 is a block cipher with the Feistel structure and consists of $68$ calls to the same round encryption routine. We used two different realizations of SIMON, the first one with a bit-parallel data-path and the second one with a bit-serial data-path \cite{aysu2014simon}.
In traditional hardware design, bit-serial methodologies are used to minimize area footprint at the expense of throughput. In the PSP execution model of software, we expect the lower gate-count of a bit-serial input specification to translate to fewer bit-wise operations in the program, and hence to a smaller code footprint. Further, we expect the bit-serial PSP design to have a lower throughput due to the lower computational effort done per synchronous clock cycle.

The first part of Table~\ref{tab:examples} shows that the models are small enough to fit on a simple embedded architecture.
Furthermore, we observe, similar to their hardware designs, the bit-serial implementation of SIMON is $20\%$ smaller than its bit-parallel counterpart in code size, whereas the bit-parallel version is $40\times$ faster and has a higher throughput than the bit-serial version. The second part of Table~\ref{tab:examples} shows the overhead of data movements. The overhead values reported are calculated as \emph{the number of move instructions (MOV, STR, LDR) divided by the total number of instructions.} Moving the data takes about 45-60\% of the entire instructions, which is expected for a straight-line program. For comparison, the data-moving overhead for a regular (non-bitsliced)  implementation of SIMON on NEON in the SUPERCOP benchmark \cite{supercopSimon} is 34\%.

\begin{figure}[t]
    \centering
    \includegraphics[width=.95\linewidth]{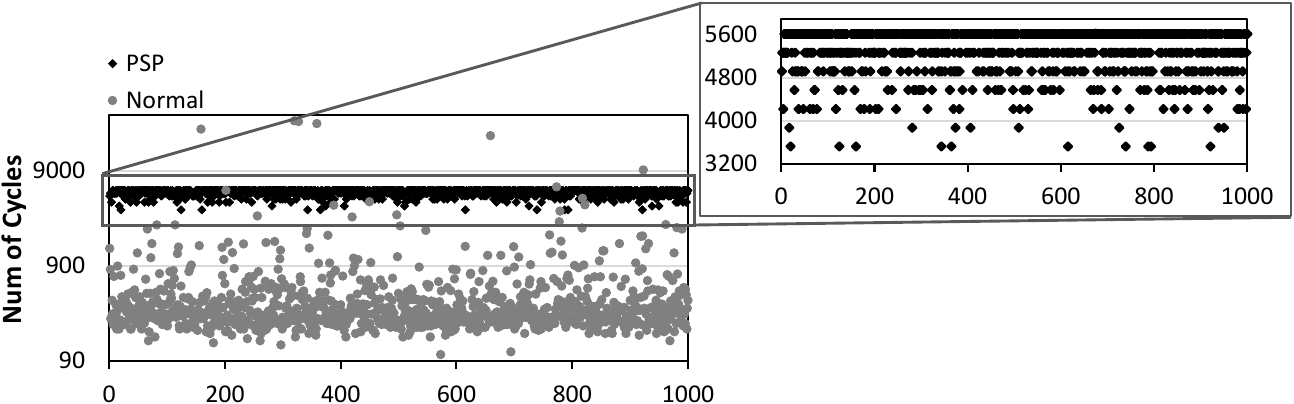}
    \vspace{-.15cm}
    \caption{Runtime of normal and PSP implementations of the GCD algorithm on 1000 random inputs.}
    \label{fig:gcdRepeatable}
    \vspace{-.8cm}
\end{figure}

We compare our PSP designs of cryptographic ciphers with their available normal implementations in Table~\ref{tab:ciphers}. In the CRYPTREC lightweight project \cite{CRYPTREC2017lightweight}, SIMON-128/128 and Midori-64 ciphers are implemented in software for the RL78 16-bit microcontroller. The throughputs of the PSP implementation of these ciphers in this work are respectively almost $1.7\times$ and $9.4\times$  higher. PRESENT-80 is evaluated in the FELICS \cite{dinu2019triathlon} project on ARM Cortex-M3. Even though the implementation of PRESENT-80 in FELICS uses pre-computed keys, still the runtime of our PSP implementation of this cipher plus its key generation is approximately $2.6\times$ smaller. 
Furthermore, to show the repeatable-timing property of PSP, we compare the runtime of the PSP and non-PSP implementations of GCD calculator for 1000 random inputs. As shown in Figure~\ref{fig:gcdRepeatable}, the PSP implementation has a quantized runtime (with steps of length the runtime of one PSP function) whereas the runtime of the normal GCD function varies with an average of 580.475 and a standard deviation of 1969.29 clock cycles.
\vspace{-.1cm}
     \vspace{-.5cm}
\section{Conclusion}\label{sec:conclusion}
We presented parallel synchronous programming as a high-throughput, fixed-time model of programming, which is beneficial in safety-critical applications. We introduced an automated method for PSP code generation that can be implemented without any dependency on commercial tools. The PSP generation can be customized for the target processor to have a better performance by defining custom libraries. Finally, through examples and discussions, we demonstrated the potential of parallel synchronous software.
\vspace{-.5cm}

%





\ifCLASSOPTIONcaptionsoff
  \newpage
\fi




\bibliographystyle{IEEEtran}
\bibliography{Bibliography}

\end{document}